\documentclass[12pt]{iopart}

\begin{document}

\title[]{Hamiltonian systems of Jordan block type: delta-functional reductions of the kinetic equation for soliton gas}

\author{P. Vergallo$^{1,2}$, E.V. Ferapontov$^{3,4}$,}

\address{$^1$ Department of Mathematical, Computer, Physical and Earth Sciences, University of Messina, V.le F. Stagno D'Alcontres 31, Messina, 98166, Italy}
\address{$^2$  Istituto Nazionale di Fisica Nucleare, Sez.\ Lecce}
\address{$^3$ Department of Mathematical Sciences, Loughborough University, Loughborough, Leicestershire LE11 3TU, United Kingdom} 
\address{$^4$ Institute of Mathematics, Ufa Federal Research Centre, Russian Academy of Sciences, 112, Chernyshevsky Street,  Ufa 450077, Russia} 
 \vspace{10pt} 
\vspace{5pt} \ead{pierandrea.vergallo@unime.it, E.V.Ferapontov@lboro.ac.uk} \vspace{10pt}

\begin{abstract}
We demonstrate that linear degeneracy is a necessary condition for  quasilinear systems of Jordan block type to possess  first-order Hamiltonian structures. Multi-Hamiltonian formulation of  linearly degenerate  systems governing delta-functional reductions of the kinetic equation for dense soliton gas is established
(for  KdV, sinh-Gordon, hard-rod, Lieb-Liniger, DNLS, and separable cases). 

\bigskip


\bigskip

\end{abstract}

{\it To our friend and colleague Maxim Pavlov on the occasion of his 60th birthday}

\section{Introduction}

In this paper we study  quasilinear systems 
 \begin{equation}\label{q}
{R}_{t}+A({R}){R}_{x}=0
\end{equation}
where ${R}=(R^1, ..., R^n)^T$ is the vector of dependent variables,  $A$ is an $n\times n$ matrix, and  $t, x$ are the independent variables. We assume that the matrix $A$ has upper-triangular Toeplitz form,
\begin{equation}\label{J}
A=\lambda^0E+\sum_{i=1}^{n-1}\lambda^iP^i;
\end{equation}
here  $E$ is the $n\times n$ identity matrix, $P$ is the $n\times n$ Jordan block with zero eigenvalue (note that $P^n=0$), and $\lambda^0, \lambda^i$ are  functions of $R$. Explicitly, a three-component version of system  (\ref{q}), (\ref{J}) is
\begin{equation} \label{T}
\left(\begin{array}{c}R^1\\ R^2\\R^3\end{array}\right)_t+
\left(\begin{array}{ccc}
\lambda^0 & \lambda^1 & \lambda^2\\
0&\lambda^0&\lambda^1\\
0&0&\lambda^0
\end{array}\right)
\left(\begin{array}{c}R^1\\ R^2\\ R^3\end{array}\right)_x=0.
\end{equation}
More generally, in what follows we will allow the matrix $A$ to be block-diagonal with several upper-triangular Toeplitz blocks of the above type. We will refer to such systems as being of Jordan block (Toeplitz block) type. 
Systems of this kind naturally arise in applications   as degenerations of hydrodynamic systems associated with multi-dimensional hypergeometric functions  \cite{KK}, in the context of parabolic regularisation of the Riemann equation \cite{KO2}, as  reductions of hydrodynamic chains and linearly degenerate dispersionless PDEs in 3D \cite{Pavlov1},  in the context of Nijenhuis geometry \cite{BKM}, and as primary flows of non-semisimple Frobenius manifolds \cite{LP}.  
It was shown in \cite{XF} that  {\it integrable} systems of  Jordan block type are governed by the modified KP hierarchy. An interesting integrable example of type (\ref{q}) where the matrix $A$ consists of several $2\times 2$ Jordan blocks, arises as a delta-functional reduction of the kinetic equation for dense soliton gas \cite{PTE, FP}.

Here we concentrate on systems (\ref{q}) that can be represented in Hamiltonian form,
$$
R^i_t+B^{ij}\frac{\delta H}{\delta R^j}=0,
$$
where $B^{ij}$ is a Hamiltonian operator of Dubrovin-Novikov type,
\begin{equation}\label{DN}
B^{ij}=g^{ij}(R)\partial_x+\Gamma^{ij}_k(R)R^k_x.
\end{equation}
The conditions for operator (\ref{DN}) to be Hamiltonian  were obtained in \cite{DN83}. In particular, if $\det(g^{ij})\neq 0$, then $g$ is a contravariant flat metric and $\Gamma^{ij}_k=-g^{is}\Gamma^{j}_{sk}$ are the contravariant  Christoffel symbols of the associated Levi-Civita connection. Thus, to specify a Hamiltonian structure of  type (\ref{DN}), it is sufficient to provide the corresponding contravariant flat metric $g^{ij}$ (as done in the examples below).
Our results can be summarised as follows.
\begin{itemize}

\item Suppose that the matrix $A$ of system (\ref{q}) consists of several upper-triangular Toeplitz blocks  (of the size  $>1$) with distinct eigenvalues. We prove that for a system of this type to be Hamiltonian, it must be linearly degenerate (Theorem 1 of Section \ref{sec:Ham}). Recall that  system (\ref{q}) is said to be linearly degenerate if the Lie derivatives of its eigenvalues along the corresponding eigenvectors are zero. We refer to Section \ref{sec:ld} for  the explicit form of conditions of linear degeneracy.

\item In Section \ref{sec:sol}  we provide multi-Hamiltonian formulation for  particularly interesting examples of  linearly degenerate systems  (\ref{q}) governing delta-functional reductions of the kinetic equation for dense soliton gas \cite{PTE, FP},
\begin{equation*}
\begin{array}{c}
f_t+(sf)_x=0,\\
s(\eta)=S(\eta)+\mathop{\int}_{0}^{\infty}G(\mu, \eta)f(\mu)[s(\mu)-s(\eta)]\ d\mu,
\end{array}
\end{equation*}
where $f(\eta)=f(\eta, x, t)$ is the distribution function and $s(\eta)=s(\eta, x, t)$ is the associated transport velocity; we refer to Section \ref{sec:sol} for further details. The  delta-functional ansatz,
\begin{equation*}
 f(\eta, x, t)=\sum_{i=1}^{n}u^i(x, t)\ \delta(\eta-\eta^i(x, t)),
\end{equation*}
leads to a quasilinear system for $u^i(x,t)$ and $\eta^i(x,t)$ whose matrix  consists of several $2\times 2$ upper-triangular Toeplitz blocks. Our analysis suggests that the requirement of existence of a Hamiltonian structure restricts the form of the 2-soliton interaction kernel $G(\mu, \eta)$:
\begin{equation}\label{G2}
G(\mu, \eta)=p(\mu)q(\eta)f[a(\mu)-b(\eta)],
\end{equation}
where $p, q, f, a, b$ are some functions of the indicated arguments. We establish the existence of local Hamiltonian structures for delta-functional reductions of KdV, sinh-Gordon, hard-rod, Lieb-Liniger, DNLS, and separable cases; note that all of them fall into class (\ref{G2}).

\end{itemize}

Let us conclude this introduction with the example of an integrable hierarchy of Jordan block type coming from the theory of associativity (WDVV) equations. The following function,
$$
F=\frac{1}{2}(u^1)^2u^3+\frac{1}{2}u^1(u^2)^2+\frac{1}{8}\frac{(u^2)^4}{u^3},
$$
has appeared in \cite{PVV}, Section 6, see also  \cite{LP}, eqn (3.18), as  a non-semisimple WDVV prepotential (we refer to the above papers for all specific WDVV-related aspects). With this function $F$ we associate two commuting systems (primary flows in the language of WDVV equations),
$$
u^1_t=(F_{u^2u^3})_x, \quad u^2_t=(F_{u^2u^2})_x, \quad u^3_t=(F_{u^1u^2})_x, 
$$
and
$$
u^1_s=(F_{u^3u^3})_x, \quad u^2_s=(F_{u^2u^3})_x, \quad u^3_s=(F_{u^1u^3})_x. 
$$
In explicit form,
$$
u^1_t=-\frac{3}{2}\frac{(u^2)^2}{(u^3)^2}u^2_x+\frac{(u^2)^3}{(u^3)^3}u^3_x, \quad u^2_t=u^1_x+3\frac{u^2}{u^3}u^2_x-\frac{3}{2}\frac{(u^2)^2}{(u^3)^2}u^3_x, \quad u^3_t=u^2_x, 
$$
and
$$
u^1_s=\frac{(u^2)^3}{(u^3)^3}u^2_x-\frac{3}{4}\frac{(u^2)^4}{(u^3)^4}u^3_x, \quad u^2_s=-\frac{3}{2}\frac{(u^2)^2}{(u^3)^2}u^2_x+\frac{(u^2)^3}{(u^3)^3}u^3_x, \quad u^3_s=u^1_x. 
$$
\medskip
Introducing the variables $R^1=-\frac{1}{u^3}, \ R^2=\frac{u^2}{u^3}, \ R^3=u^1+\frac{1}{2}\frac{(u^2)^2}{u^3}$, we can rewrite these commuting systems in the upper-triangular Toeplitz form,
\begin{equation}\label{E1}
\left(\begin{array}{c}R^1\\ R^2\\R^3\end{array}\right)_t=
\left(\begin{array}{ccc}
R^2 &-R^1 & 0\\
0&R^2&-R^1\\
0&0&R^2
\end{array}\right)
\left(\begin{array}{c}R^1\\ R^2\\ R^3\end{array}\right)_x
\end{equation}
and 
\begin{equation}\label{sys2}
\left(\begin{array}{c}R^1\\ R^2\\R^3\end{array}\right)_s=
\left(\begin{array}{ccc}
-\frac{1}{2}(R^2)^2 &R^1R^2 & (R^1)^2\\
0&-\frac{1}{2}(R^2)^2 &R^1R^2\\
0&0&-\frac{1}{2}(R^2)^2 
\end{array}\right)
\left(\begin{array}{c}R^1\\ R^2\\ R^3\end{array}\right)_x,
\end{equation}
respectively. A general commuting flow of the hierarchy generated by systems (\ref{E1}), (\ref{sys2}) has the form
\[
\left(\begin{array}{c}R^1\\ R^2\\R^3\end{array}\right)_{\tau}=
\left(\begin{array}{ccc}
\mu^0 &\mu^1 & \mu^2\\
0&\mu^0 &\mu^1\\
0&0&\mu^0 
\end{array}\right)
\left(\begin{array}{c}R^1\\ R^2\\ R^3\end{array}\right)_x;
\]
here $\mu^0=f, \ \mu^1=-R^1f_{R^2}, \ \mu^2=-(R^1)^2f_{R^2R^2}-R^1f_{R^3}$ where $f(R^2, R^3)=p(R^3)+q(R^3)R^2+r(R^3)(R^2)^2$ is a quadratic polynomial in $R^2$, and $f_{R^2}, \ f_{R^3}$, etc, indicate partial derivatives. It follows from \cite{PVV, LP} that this hierarchy is bi-Hamiltonian. 
In fact, our calculations demonstrate that it possesses infinitely many compatible Hamiltonian structures (\ref{DN})
with the flat contravariant metrics
\begin{equation}{\footnotesize 
g^{ij}=(R^1)^2\left(\begin{array}{ccc}
(-s_1'+2R^2s_2+2s_3)R^1&\frac{(R^2)^2}{2}s_2+R^2s_3+s_4&s_1\\
\frac{(R^2)^2}{2}s_2+R^2s_3+s_4&s_1&0\\
s_1&0&0
\end{array}\right)},
\end{equation}
where $s_1, s_2, s_3, s_4$ are arbitrary functions of the variable $R^3$.

\section{Linearly degenerate  systems of Jordan block type}
\label{sec:ld}

Recall that a strictly hyperbolic quasilinear system  is said to be {\it linearly degenerate} if its eigenvalues (characteristic speeds) are constant in the direction of the corresponding  eigenvectors. Explicitly, $L_{r^i}\lambda^i=0$, no summation, where $L_{r^i}$ is  the Lie derivative of the eigenvalue $\lambda^i$ in the direction of the corresponding eigenvector $r^i$. Linearly degenerate systems are quite exceptional from the point of view of solvability of the initial value problem; they have been thoroughly investigated in the literature, see e.g. \cite{R1, Liu, Serre}. There exists a simple invariant criterion of linear degeneracy which does not appeal to eigenvalues/eigenvectors. 
Let us introduce the characteristic polynomial of $A$,
 $$
 P(\lambda)=det(\lambda E-A)={\lambda  }^n  +  f_1({R}){\lambda
}^{n-1} +f_2({R}){\lambda}^{n-2}+ \ldots + f_n({R}).
$$
The condition of linear degeneracy can be represented in the form \cite{Fer},
\begin{equation}\label{lindeg}
\nabla f_1~A^{n-1}+\nabla  f_2~A^{n-2}+\ldots  +\nabla  f_n=0,
\end{equation}
where   $\nabla f=({{\partial
f}\over {\partial R^1}},\ldots , {{\partial f}\over {\partial
R^n}})$ is the gradient, and $A^k$ denotes  $k$-th power of the matrix $A$. Equivalently, one can write (\ref{lindeg}) in the form
\begin{equation}\label{lindeg1}
\nabla P(\lambda)\vert_{\lambda=A}=0.
\end{equation}
Note that condition (\ref{lindeg}) can be seen as a {\it definition} of linear degeneracy for arbitrary quasilinear systems, not necessarily strictly hyperbolic. 

\medskip

\noindent{\bf Proposition 1.} {\it For systems (\ref{q}), (\ref{J}) of Jordan block type, the condition of linear degeneracy is equivalent to }
\begin{equation}\label{l}
\frac{\partial \lambda^0}{\partial R^1}=0.
\end{equation}
\medskip

\centerline{\bf Proof:}
\medskip

For systems (\ref{q}), (\ref{J}), the characteristic polynomial takes the form 
 $$
 P(\lambda)=det(\lambda E-A)=(\lambda-\lambda^0)^n,
$$
and the condition of linear degeneracy (\ref{lindeg1}) reduces to
$$
0=\nabla P(\lambda)\vert_{\lambda=A}=-n\nabla \lambda_0(\lambda-\lambda^0)^{n-1}\vert_{\lambda=A}
=-n\nabla \lambda_0(A-\lambda_0E)^{n-1}.
$$
It remains to note  that the matrix $(A-\lambda_0E)^{n-1}$ has only one nonzero element, namely 
$(\lambda^1)^{n-1}$, in the upper right corner (since  $A$ is a single Jordan block, the coefficient $\lambda^1$ must be non-zero). This ends the proof.

\medskip

\noindent{\bf Remark.} Note that, in full analogy with the strictly hyperbolic case, condition (\ref{l}) is equivalent to the requirement that the Lie derivative of the eigenvalue $\lambda^0$ in the direction of the corresponding (unique) eigenvector $\frac{\partial}{\partial R^1}$ vanishes.
\medskip

Proposition 1 extends to the general case where the matrix $A$ consists of several upper-triangular Toeplitz blocks with distinct eigenvalues.

\medskip

\noindent{\bf Proposition 2.} {\it Suppose that the matrix $A$ has block-diagonal form with several  blocks  $J_{\alpha}$ of type (\ref{J}) with distinct eigenvalues $\lambda^0_{\alpha}$. Then the condition of linear degeneracy is equivalent to 
\begin{equation}\label{l1}
\frac{\partial \lambda^0_{\alpha}}{\partial R^1_{\alpha}}=0 \qquad \forall \alpha, 
\end{equation}
no summation}.
\medskip

\centerline{\bf Proof:}
\medskip

We will outline the proof in the case of two Jordan blocks; the general case is analogous. For two Jordan blocks of size $m_1\times m_1$ and $m_2\times m_2$ with eigenvalues $\lambda^0_1$ and $\lambda^0_2$, the characteristic polynomial takes the form 
 $$
 P(\lambda)=det(\lambda E-A)=(\lambda-\lambda^0_1)^{m_1}(\lambda-\lambda^0_2)^{m_2},
$$
and the condition of linear degeneracy (\ref{lindeg1}) reduces to
$$
0=\nabla P(\lambda)\vert_{\lambda=A}
$$
$$
=-m_1\nabla \lambda_0^1(\lambda-\lambda^0_1)^{m_1-1}(\lambda-\lambda^0_2)^{m_2}\vert_{\lambda=A}
-m_2\nabla \lambda_0^2(\lambda-\lambda^0_1)^{m_1}(\lambda-\lambda^0_2)^{m_2-1}\vert_{\lambda=A}
$$
$$
=-m_1\nabla \lambda_0^1(A-\lambda^0_1E)^{m_1-1}(A-\lambda^0_2E)^{m_2}
-m_2\nabla \lambda_0^2(A-\lambda^0_1E)^{m_1}(A-\lambda^0_2E)^{m_2-1}.
$$
It remains to note that the matrix $(A-\lambda^0_1E)^{m_1-1}(A-\lambda^0_2E)^{m_2}$ has only one nonzero entry, namely, $(\lambda^1_1)^{m_1-1}(\lambda^0_1-\lambda^0_2)^{m_2}$, in the upper right corner of the $m_1\times m_1$ diagonal block. 
Similarly, the matrix $(A-\lambda^0_1E)^{m_1}(A-\lambda^0_2E)^{m_2-1}$ has only one nonzero entry, namely, $(\lambda^1_2)^{m_2-1}(\lambda^0_2-\lambda^0_1)^{m_1}$, in the upper right corner of the $m_2\times m_2$ diagonal block. 
This ends the proof.

\section{Linear degeneracy of Hamiltonian  systems of Jordan block type}
\label{sec:Ham}

The main result of this section is as follows.

\bigskip

\noindent{\bf Theorem 1.} {\it Suppose that the matrix $A$ of system (\ref{q}) has block-diagonal form with several blocks $J_{\alpha}$ of  type (\ref{J}) (of the size $n_{\alpha}\times n_{\alpha}$ with $n_{\alpha}>1$) having distinct eigenvalues. Then the existence of Hamiltonian structure (\ref{DN})  implies linear degeneracy.}

\medskip

\centerline{\bf Proof:}
\medskip

It was shown by Tsarev   \cite{Tsarev, Tsarev1} that system (\ref{q}) admits Hamiltonian formulation (\ref{DN})
if and only if the following conditions are satisfied:
\begin{equation}
g^{is}A^j_{s}=g^{js}A^i_s,
\label{cond1}
\end{equation}
\begin{equation}
\label{cond2} 
\nabla_iA^j_k=\nabla_kA^j_i,
\end{equation}
where  $A=(A^i_j)$ is the matrix of the system and $\nabla$ denotes covariant derivative in the Levi-Civita connection of the metric $g$. Note that conditions (\ref{cond1}) imply that, if $A$ has block-diagonal form with several upper-triangular Toeplitz blocks $J_{\alpha}$  having  distinct eigenvalues, then the metric $g$ (with low indices) also has block-diagonal form with (low-triangular) Hankel blocks $g_{\alpha}$  of the same size. Here is the form of $A$ and $g$ in the case of a single block of size  $3\times 3$:
\[
A=
\left(\begin{array}{ccc}
\lambda^0 & \lambda^1 & \lambda^2\\
0&\lambda^0&\lambda^1\\
0&0&\lambda^0
\end{array}\right),
\qquad
g=
\left(\begin{array}{ccc}
0 & 0 & r\\
0&r&q\\
r&q&p
\end{array}\right).
\]
For definiteness, let us assume that $A$ is an $n\times n$ matrix composed of two Toeplitz blocks of the size $m_1\times m_1$ and $m_2\times m_2$, with distinct eigenvalues $\lambda^0_1\neq \lambda^0_2$ (the general case is analogous). Let us suppose that the system has Hamiltonian structure (\ref{DN}), so that Tsarev's conditions are satisfied. 
Since $g$ (with low indices) consists of two (low-triangular) Hankel blocks, it follows  that $g_{1s}=0$ for every  $s\neq m_1$ and $g_{m_1+1,s}=0$ for every $s\neq n$. As the inverse of a low-triangular Hankel matrix is an upper-triangular Hankel matrix, for the metric with upper indices we obtain  $g^{m_1s}=0$ for every $s\neq 1$ and $g^{ns}=0$ for every $s\neq m_1+1$. By this, we obtain that 
$\Gamma^{m_1}_{1s}=\Gamma^n_{m_1+1,s}=0$ for every $s$. Keeping in mind that $m_1>1$ and $m_2>1$ and using the notation for the dependent variables
$$
(R^1, \dots, R^n)=(R^1_1, \dots, R^{m_1}_1, R^1_2, \dots, R^{m_2}_2),
$$
by (\ref{cond2}) we obtain (no summation on the repeated index $m_1$):
\begin{equation*}
\begin{array}{c}
\displaystyle 0=\nabla_1A^{m_1}_{m_1}-\nabla_{m_1}A^{m_1}_1=\frac{\partial A^{m_1}_{m_1}}{\partial R^1}-\frac{\partial A^{m_1}_{1}}{\partial R^{m_1}}+\Gamma^{m_1}_{1s}A^s_{m_1}-\Gamma^{m_1}_{m_1s}A^s_1\\
\ \\
\displaystyle =\frac{\partial A^{m_1}_{m_1}}{\partial R^1}=\frac{\partial \lambda^0_1}{\partial R^1_1}.
\end{array}
\end{equation*}
Analogously (no summation on the index $n$),
\begin{equation*}
0=\nabla_{m_1+1}A^{n}_{n}-\nabla_{n}A^{n}_{m_1+1}=\frac{\partial A^{n}_{n}}{\partial R^{m_1+1}}=\frac{\partial \lambda^0_2}{\partial R^1_2}.
\end{equation*}
By Proposition 2, the statement is proved. 

\medskip

\noindent {\bf Remark.} A class of integrable systems of Jordan block type (that are not linearly degenerate) was described in \cite{XF} in terms of the modified KP hierarchy. By the above Theorem, none of these systems are Hamiltonian  in the Dubrovin-Novikov sense.

\section{Kinetic equation for soliton gas: reductions of Jordan block type}
\label{sec:sol}

Our interest in systems of Jordan block type stems from the study of El's integro-differential kinetic equation for  dense soliton gas \cite{El,EK, EKPZ}:
\begin{equation}\label{gas}
\begin{array}{c}
f_t+(sf)_x=0,\\
\ \\
s(\eta)=S(\eta)+\mathop{\int}_{0}^{\infty}G(\mu, \eta)f(\mu)[s(\mu)-s(\eta)]\ d\mu,
\end{array}
\end{equation}
where $f(\eta)=f(\eta, x, t)$ is the distribution function and $s(\eta)=s(\eta, x, t)$ is the associated transport velocity. Here the variable $\eta$ is a spectral parameter in the Lax pair associated with the dispersive hydrodynamics; the function $S(\eta)$ (free soliton velocity) and the  kernel $G(\mu, \eta)$ (symmetrised phase shift due to pairwise soliton collisions)  are independent of $x$ and $t$. 
The kernel $G(\mu, \eta)$ is assumed to be symmetric: $G(\mu, \eta)=G(\eta, \mu)$. Equation (\ref{gas}) describes the evolution of a dense soliton gas and represents a broad generalisation of Zakharov's kinetic equation for rarefied soliton gas \cite{Z}. It has appeared independently in the context of generalised hydrodynamics of multi-body quantum integrable systems \cite{Do}.
 In the special case 
$$
S(\eta)=4\eta^2, \qquad G(\mu, \eta)=\frac{1}{\eta \mu} \log \bigg|{ \frac{\eta-\mu}{\eta+\mu}}\bigg|,
$$
system (\ref{gas}) was derived in \cite{El} as a thermodynamic limit of the KdV Whitham equations. It was demonstrated in \cite{PTE} that under a delta-functional ansatz,
\begin{equation}\label{del}
 f(\eta, x, t)=\sum_{i=1}^{n}u^i(x, t)\ \delta(\eta-\eta^i(x, t)),
\end{equation}
system (\ref{gas}) reduces to a $2n\times 2n$ quasilinear system for $u^i(x, t)$ and $\eta^i(x, t)$,
\begin{equation}\label{uv}
u^i_t=(u^iv^i)_x, \qquad \eta^i_t=v^i\eta^i_x,
\end{equation} 
where $v^i\equiv -s(\eta^i, x, t)$ can be recovered from the linear system
\begin{equation}\label{vlin}
v^i=-S(\eta^i)+\sum_{k\ne i}\epsilon^{ki}u^k(v^k-v^i), \qquad  \epsilon^{ki}={G(\eta^k, \eta^i)}, \ k\ne i.
\end{equation}
The special choice $\eta^i(x, t)=const$ was discussed  in \cite{EKPZ}. In this case, the last $n$ equations (\ref{uv}) are satisfied identically, while the first $n$ equations constitute an integrable diagonalisable linearly degenerate system whose Hamiltonian aspects were explored in \cite{DZP}. The case of non-constant $\eta^i(x, t)$ was investigated recently  in \cite{FP}; in this case the matrix of the corresponding system (\ref{uv}) is reducible to $n$ Jordan blocks of size $2\times 2$, furthermore, it was shown that the system is integrable by a suitable extension of the generalised hodograph method of \cite{Tsarev, Tsarev1}. 
Following \cite{PTE}, let us introduce the new  variables $r^i$ by the formula
$$
r^i=-\frac{1}{u^i}\left(1+\sum_{k\ne i}\epsilon^{ki}u^k \right).
$$
In the dependent variables $r^i, \eta^i$, system (\ref{uv}) reduces to block-diagonal form 
\begin{eqnarray}
\label{J1}
&&
\begin{array}{l} 
r^i_t=v^i r^i_x+p^i\eta^i_x,  \\
\eta^i_t=v^i\eta^i_x,  
\end{array}
\end{eqnarray}
$i=1, \dots, n$, which consists of $n$ Jordan blocks of size $2\times 2$. Here the coefficients $v^i$ and $p^i$ can be expressed in terms of $(r, \eta)-$variables as follows. Let us introduce the $n\times n$ matrix $\hat \epsilon$ with diagonal entries $r^1, \dots, r^n$ (so that $\epsilon^{ii}=r^i$) and off-diagonal entries $\epsilon^{ik}={G(\eta^i, \eta^k)}, \ k\ne i$. 
 Note that this matrix is symmetric due to the symmetry of the kernel $G$. 
 Define another symmetric matrix $\hat \beta=-\hat \epsilon^{-1}$. Explicitly, for $n=2$ we have
 $$
 \hat \epsilon=\left(
 \begin{array}{cc}
 r^1&\epsilon^{12}\\
 \epsilon^{12}&r^2
 \end{array}
 \right), \qquad
  \hat \beta=\frac{1}{r^1r^2-(\epsilon^{12})^2}\left(
 \begin{array}{cc}
 -r^2&\epsilon^{12}\\
 \epsilon^{12}&-r^1
 \end{array}
 \right).
 $$
Denote $\beta_{ik}$ the matrix elements of $\hat \beta$ (indices $i$ and $k$ are allowed to coincide). Introducing the notation $\xi^k(\eta^k)=-S(\eta^k)$, we have  the following formulae for $u^i, v^i$ and $p^i$ \cite{PTE}:
\begin{equation}\label{vp}
u^i=\sum_{k=1}^n\beta_{ki}, \ \  v^i=\frac{1}{u^i}\sum_{k=1}^n \beta_{ki}\xi^k, \ \
p^i=\frac{1}{u^i}\left(\sum_{k=1}^n \epsilon^{ki}_{,\eta^i}(v^k-v^i)u^k+(\xi^i)' \right)
\end{equation}
where we use the notation $\epsilon^{ki}_{,\eta^i}$ to indicate partial derivative with respect to  $\eta^i$.

In what follows, we investigate Hamiltonian aspects of equations (\ref{J1}), with an emphasis on the simplest nontrivial case  $n=2$. We establish the existence of local Hamiltonian structures for all standard examples such as:
\medskip

{\noindent} {\bf KdV soliton gas:}
\begin{equation*}
S(\eta)=4\eta^2, \qquad \quad\, G(\mu, \eta)=\frac{1}{\eta \mu} \log \bigg|{ \frac{\eta-\mu}{\eta+\mu}}\bigg|.
\end{equation*}
{\noindent} {\bf Sinh-Gordon soliton gas:}
\begin{equation*}
S(\eta)=\tanh \eta, \qquad  G(\mu, \eta)=\frac{1}{\cosh \eta \cosh  \mu}\  \frac{g^2\cosh(\eta-\mu)}{4\sinh^2(\eta-\mu)}.
\end{equation*}
{\noindent} {\bf Hard-rod gas:}
\begin{equation*}
S(\eta)=\eta, \qquad \qquad G(\mu, \eta)=-a.
\end{equation*}
{\noindent} {\bf Lieb-Liniger gas:}
\begin{equation*}
S(\eta)=\eta, \qquad \qquad G(\mu, \eta)=\frac{2g}{g^2+(\eta -\mu)^2}.
\end{equation*}
{\noindent} {\bf DNLS soliton gas:}
\begin{equation*}
S(\eta)=\eta, \qquad  G(\eta, \mu)=\frac{1}{2\sqrt{\eta^2-1}\sqrt{\mu^2-1}}\log  \left( \frac{(\eta-\mu)^2-\left(\sqrt{\eta^2-1}+\sqrt{\mu^2-1}\right)^2}{(\eta-\mu)^2-\left(\sqrt{\eta^2-1}-\sqrt{\mu^2-1}\right)^2}\right), 
\end{equation*}

{\noindent} {\bf Separable case:}
\begin{equation*}
S(\eta)\ {\rm arbitrary}, \qquad G(\mu, \eta)=\phi(\eta)+\phi(\mu).
\end{equation*}
We refer to \cite{El, EK, CER, Do, Do1, Do2, Spohn} for further discussion and references.

\subsection{Hamiltonian formulation of two-component reductions ($n=1$)}
\label{4.1}

In this case system (\ref{J1}) is a single $2\times 2$ Jordan block,
\begin{eqnarray*}
&&
\begin{array}{l} 
r_t=\xi\, r_x-r\xi'\, \eta_x,  \\
\eta_t=\xi\, \eta_x,  
\end{array}
\end{eqnarray*}
where $\xi(\eta)$ is some function of the indicated argument; note that the phase shift $\epsilon$ does not enter the system for $n=1$. This system possesses infinitely many Hamiltonian structures with the flat contravariant metric
$$
g^{ij}=\left(\begin{array}{cc}
f_1r^4+f_2 r^3&f_3\, r^2\\
f_3\, r^2 &0
\end{array}\right)
$$
where $f_1,f_2,f_3$ are arbitrary functions of the variable $\eta$ (here superscripts of $r$ denote powers of $r$, not indices; this convention applies to section \ref{4.1} only). The corresponding Hamiltonian operator $B^{ij}$ has the form
\begin{equation*}
B^{ij}=
\left(\begin{array}{cc}
f_1 r^4+f_2 r^3&f_3\, r^2\\
f_3\, r^2 &0
\end{array}\right)\partial_x
\end{equation*}
\begin{equation*}
\hphantom{ciaoocio}
+\left(\begin{array}{cc}
(2f_1\, r^3+\frac{3}{2}f_2\, r^2)\, r_x+\frac{1}{2}\left(f'_1\, r^4+f_2'\, r^3\right)\, \eta_x&\left(\frac{1}{2}f_2+f_3'\right)r^2\, \eta_x\\
(3f_3\, r)\, r_x-\frac{1}{2}f_2\, r^2\, \eta_x&0
\end{array}\right).
\end{equation*}

\subsection{Hamiltonian formulation of four-component  reductions ($n=2$)}

The corresponding  $4\times 4$ system (\ref{J1}) has two Jordan blocks; setting $\epsilon^{12}\equiv \epsilon$ in (\ref{vp}) we have
$$
v^1=\frac{r^2\xi^1-\epsilon \xi^2}{r^2-\epsilon}, \qquad v^2=\frac{r^1\xi^2-\epsilon \xi^1}{r^1-\epsilon},
$$
and
$$
p^1=\frac{\epsilon^2-r^1r^2}{r^2-\epsilon}\left(\frac{\xi^1-\xi^2}{r^2-\epsilon}\epsilon_{,\eta^1}+(\xi^1)'  \right), \qquad
p^2=\frac{\epsilon^2-r^1r^2}{r^1-\epsilon}\left(\frac{\xi^2-\xi^1}{r^1-\epsilon}\epsilon_{,\eta^2}+(\xi^2)'  \right),
$$
where $\xi^1=\xi^1(\eta^1), \ \xi^2=\xi^2(\eta^2)$ and $\epsilon(\eta^1, \eta^2)$ are functions of the indicated arguments. Direct computation of Tsarev's conditions (\ref{cond1}), (\ref{cond2}) can be summarised as follows.
First of all, condition (\ref{cond1}) implies that the metric (with upper indices) has block-diagonal Hankel form, 
\begin{equation}\label{g2}
g^{ij}=\left(\begin{array}{cccc}
m_1 & n_1 & 0 &0\\
n_1&0&0 &0\\
0&0&m_2& n_2\\
0&0&n_2&0
\end{array}\right),
\end{equation}
while condition (\ref{cond2}) specifies the form of $m_i, n_i$ as 
\begin{equation}\label{pq}
\begin{array}{c}
\displaystyle m_1=  \frac { \left( -2{s_1} \,  \left( {r^1}-\epsilon  \right) 
\epsilon_{,\eta^1} +\,{g_1}\, \left( r^2-\epsilon \right)  \right)  \left( {r^1r^2}- \epsilon^{2} \right) ^{2}}{ \left( r^2-\epsilon   \right) ^{3}},\\
\ \\
\displaystyle n_1=  \frac {{s_1}\,\left({r^1r^2}-\epsilon^2 \right)^{2}}{ \left( r^2-\epsilon \right) ^{2}},\\
\ \\
\displaystyle m_2=  \frac { \left( -
2\,{ s_2} \,  \left( {r^2}-\epsilon
  \right)\epsilon_{,{\eta^2}}  +{g_2}\, \left( {r^1}-\epsilon
   \right)  \right)  \left( {r^1}\,{
r^2}   
-\epsilon ^{2}\right) ^{2}}{ \left( {r^1}-\epsilon \right) ^{3}},\\
\ \\
\displaystyle n_2=  \frac {{s_2}\,
 \left( r^1r^2- \epsilon^{2
} \right) ^{2}}{ \left(r^1- \epsilon\right) ^{2}},
\end{array}
\end{equation}
where $s_1(\eta^1),\, s_2(\eta^2)$ and $g_1(r^1,\eta^1),\, g_2(r^2,\eta^2)$ are some  functions of the indicated arguments (to be determined from the flatness conditions). Note that $ r^1r^2- \epsilon^{2}=\det \hat \epsilon$.
The  analysis of flatness conditions depends on the  explicit form of the 2-soliton interaction kernel
$\epsilon(\eta^1, \eta^2)$ entering the system, and is summarised on a case-by-case basis (we only present the corresponding functions $s_1(\eta^1),\, s_2(\eta^2)$ and $g_1(r^1,\eta^1),\, g_2(r^2,\eta^2)$ from (\ref{pq}), as well as the corresponding Hamiltonian densities). We emphasise that all examples discussed below are multi-Hamiltonian, possessing two or more compatible Hamiltonian structures.

\bigskip

{\noindent} {\bf KdV soliton gas:}
$$
\xi^i(\eta^i)=4(\eta^i)^2, \qquad \epsilon(\eta^1, \eta^2)=\frac{1}{\eta^1 \eta^2} \log \bigg|{ \frac{\eta^1-\eta^2}{\eta^1+\eta^2}}\bigg|.
$$
The requirement of  flatness  of metric (\ref{g2}) leads to the following expressions for the functions $s_1(\eta^1), \ s_2(\eta^2)$ and $g_1(r^1, \eta^1), \ g_2(r^2, \eta^2)$:
$$
s_1=-\frac{(c_1+c_2)\eta^1}{4},\quad s_2=-\frac{(c_1+c_2)\eta^2}{4},
$$
$$
g_1=c_1r^1,\quad g_2=c_2r^2,
$$
where $c_1,c_2$ are arbitrary constants. Based on the general form of  conservation laws from  \cite{FP}, we obtain the corresponding Hamiltonian density:

\medskip 

\begin{equation*}
h=-8\frac{(c_1+2c_2)(\epsilon-r^2)(\eta^1)^2+(c_2+2c_1)(\epsilon-r^1)(\eta^2)^2}{(c_1+2c_2)(c_2+2c_1)(r^1r^2-\epsilon^2)}.
\end{equation*}
Note that only one structure from this two-parameter family generalises to arbitrary $n$, namely, the one with $c_1=c_2=c$ (without any loss of generality, we will set $c=1$). The corresponding flat metric (with upper indices) of Hamiltonian operator (\ref{DN}) consists of $n$ upper-triangular Toeplitz blocks, 
$$
\left(\begin{array}{cc}
m_i & n_i \\
n_i&0\\
\end{array}\right),
$$
where
\begin{equation*}
n_i=-\frac{\eta^i}{2(u^i)^2}, \quad m_i=\frac{\eta^i}{(u^i)^3}\displaystyle \sum_{j\neq i}^n u^j\epsilon^{ji}_{,\eta^i} +\frac{r^i}{(u^i)^2},
\end{equation*}
and the Hamiltonian  density is given by
\begin{equation*}
h=-\frac{8}{3}\displaystyle \sum_{i=1}^N u^i(\eta^i)^2;
\end{equation*}
here the variables $u^i$ are defined by formula (\ref{vp}).

\bigskip

{\noindent} {\bf Sinh-Gordon soliton gas:}
$$
\xi^i(\eta^i)=\tanh \eta^i, \qquad \epsilon(\eta^1, \eta^2)=\frac{1}{\cosh \eta^1 \cosh  \eta^2}\  \frac{g^2\cosh(\eta^1-\eta^2)}{4\sinh^2(\eta^1-\eta^2)}.
$$
The requirement of  flatness  of metric (\ref{g2}) leads to the following expressions for the functions $s_1(\eta^1), \ s_2(\eta^2)$ and $g_1(r^1, \eta^1), \ g_2(r^2, \eta^2)$:
$$
s_1=-\frac{c_2}{2}, \qquad s_2=-\frac{c_2}{2}, 
$$
$$
g_1=(c_1+c_2\tanh{(\eta^1)})r^1, 
\qquad g_2=(-c_1+c_2\tanh{(\eta^2)})r^2,
$$
where $c_1,c_2$ are arbitrary constants.  The corresponding Hamiltonian density is given by 
\begin{equation*}
h=\frac{(\epsilon-r^2)\psi^1+(\epsilon-r^1)\psi^2}{r^1r^2-\epsilon^2}
\end{equation*}
where 
\begin{equation*}
\psi^1=-2\frac{\displaystyle\int{\left(e^{\frac{\eta^1(c_1-c_2)}{c_2}}+e^{\frac{\eta^1(c_1+c_2)}{c_2}}\right)\tanh{\eta^1}\,  d\eta^1}}{c_2(1+e^{-2\eta^1})}e^{-\frac{\eta^1(c_1+c_2)}{c_2}},
\end{equation*}
\begin{equation*}
\psi^2=-2\frac{\displaystyle\int{\left(e^{-\frac{\eta^2(c_1-c_2)}{c_2}}+e^{-\frac{\eta^2(c_1+c_2)}{c_2}}\right)\tanh{\eta^2}\,  d\eta^2}}{c_2(1+e^{2\eta^2})}e^{\frac{\eta^2(c_1+c_2)}{c_2}}.
\end{equation*}
\bigskip

\noindent {\bf Hard rod gas:} 
$$
\xi^i(\eta^i)=\eta^i, \qquad \epsilon(\eta^1, \eta^2)=-a=const.
$$
The requirement of  flatness  of metric (\ref{g2}) implies that the  functions $s_1(\eta^1), \ s_2(\eta^2)$ in (\ref{pq})  remain arbitrary, while $g_1(r^1, \eta^1), \ g_2(r^2, \eta^2)$  specialise to
$$
g_1=c_1\, (r^1)^2- 2c_3r^1- c_2\, a^2, \qquad
g_2=c_2\, (r^2)^2+2c_3r^2-c_1\, a^2, 
$$
where  $c_1,c_2,c_3$ are arbitrary constants. Thus, we have an infinity of local compatible Hamiltonian structures parametrised by two arbitrary functions of one variable and three arbitrary constants.
The coefficients of the corresponding contravariant metric  (\ref{g2}) take the form
\begin{equation*}
m_1=\frac{\left(c_1\, (r^1)^2-2c_3r^1-c_2\, a^2\right)\, (a^2-r^1r^2)^2}{(a+r^2)^2},
\end{equation*}
\begin{equation*}
n_1=\frac{s_1\, (a^2-r^1r^2)^2}{(a+r^2)^2},
\end{equation*}
\begin{equation*}
m_2=\frac{\left(c_2\, (r^2)^2+2c_3r^2-c_1\, a^2\right)\, (a^2-r^1r^2)^2}{(a+r^1)^2},
\end{equation*}
\begin{equation*}
n_2=\frac{s_2\, (a^2-r^1r^2)^2}{(a+r^1)^2}.
\end{equation*}
The corresponding Hamiltonian density is  given by
\begin{equation*}
h=\frac{(a+r^2)\psi^1+(a+r^1)\psi^2}{r^1r^2-a^2}+\sigma^1(\eta^1)+\sigma^2(\eta^2)
\end{equation*}
where 
$$
\psi^1=c_2k\left(ac_1s_2^2\sigma_2''-c_3s_1^2\sigma_1''-s_1\left(s_1'c_3-c_1c_2a^2-{c_3}^2\right)\sigma_1'+c_1\left(as_2s_2'\sigma_2'+a\xi^2c_2-\xi^1c_3\right)\right),
$$
$$
\psi^2=c_1k\left(ac_2s_1^2\sigma''_1+c_3s_2^2\sigma_2''+s_2\left(s_2'c_3+c_1c_2a^2+c_3^2\right)\sigma'_2+c_2\left(as_1s_1'\sigma_1'+a\xi^1c_1+\xi^2c_3\right)\right).
$$
Here 
 $$
k=\frac{1}{c_1c_2(c_1c_2a^2+c_3^2)},
$$
and   $\sigma_1(\eta^1), \sigma_2(\eta^2)$ are solutions of the following ODEs:
\begin{equation*}
(s_1^2)\, \sigma_1'''+(3s_1's_1)\, \sigma_1''+\left(-c_1c_2a^2-c_3^2+(s_1')^2+s_1s_1''\right)\, \sigma_1'+c_1(\xi^1)'=0,
\end{equation*}
\begin{equation*}
(s_2^2)\, \sigma_2'''+(3s_2's_2)\, \sigma_2''+\left(-c_1c_2a^2-c_3^2+(s_2')^2+s_2s_2''\right)\, \sigma_2'+c_2(\xi^2)'=0.
\end{equation*}

\bigskip

The  formulae for contravariant metric for the hard rod gas generalise to the case of arbitrary $n>2$ in  the obvious way:
$$
m_i=\frac{\left({c}_ir^i-a\sum_{k\ne i}c_k\right)(a+r^i)}{\prod_{k\ne i}(a+r^k)^2}(\det{\hat{\epsilon}})^2, \qquad
n_i=\frac{s_i(\eta^i)}{\prod_{k\ne i}(a+r^k)^2}(\det{\hat{\epsilon}})^2.
$$
Note that although this formula gives {\it all} Hamiltonian structures for the $n> 2$ hard rod gas, it gives only a subfamily thereof for $n=2$, namely those for which $2c_3+a(c_1-c_2)=0$. 

\bigskip

{\noindent} {\bf Lieb-Liniger gas:}
$$
\xi^i(\eta^i)=\eta^i, \qquad \epsilon(\eta^1, \eta^2)=\frac{2a}{a^2+(\eta^1 -\eta^2)^2}.
$$
The requirement of  flatness  of metric (\ref{g2}) leads to the following expressions for the functions $s_1(\eta^1), \ s_2(\eta^2)$ and $g_1(r^1, \eta^1), \ g_2(r^2, \eta^2)$:
$$
s_1=c_2,\quad s_2=c_2,
$$
$$g_1=-c_1r^1,\quad g_2=c_1r^2,
$$
where $c_1, c_2$ are arbitrary constants.
The corresponding Hamiltonian density is given by
\begin{equation*}
h=\frac{(2c_1\eta^1-4c_2)(\epsilon-r^2)-(2c_1\eta^2+4c_2)(\epsilon-r^1)}{c_1^2(r^1r^2-\epsilon^2)}.
\end{equation*}
\bigskip

{\noindent} {\bf DNLS soliton gas:}{\footnotesize
$$
\xi^i(\eta^i)=\eta^i, \quad {\footnotesize\epsilon(\eta^1, \eta^2)=\frac{1}{2\sqrt{(\eta^1)^2-1}\sqrt{(\eta^2)^2-1}}\log  \left( \frac{(\eta^1-\eta^2)^2-\left(\sqrt{(\eta^1)^2-1}+\sqrt{(\eta^2)^2-1}\right)^2}{(\eta^1-\eta^2)^2-\left(\sqrt{(\eta^1)^2-1}-\sqrt{(\eta^2)^2-1}\right)^2}\right). }
$$}
This expression can be equivalently written in form (\ref{G2}),
$$
\epsilon(\eta^1, \eta^2)=\frac{1}{\sqrt{(\eta^1)^2-1}\sqrt{(\eta^2)^2-1}}\log \coth \left( \frac{1}{4}\log \frac{\eta^1+1}{\eta^1-1}- \frac{1}{4}\log \frac{\eta^2+1}{\eta^2-1}\right).
$$
The requirement of  flatness  of metric (\ref{g2}) leads to the following expressions for the functions $s_1(\eta^1), \ s_2(\eta^2)$ and $g_1(r^1, \eta^1), \ g_2(r^2, \eta^2)$:
$$
s_1=-\frac{c_1}{2}((\eta^1)^2-1),\quad s_2=-\frac{c_1}{2}((\eta^2)^2-1),
$$
$$g_1=(c_1\eta^1+c_2)r^1,\quad g_2=(c_1\eta^2-c_2)r^2.
$$
The Hamiltonian density is  given by
\begin{equation*}
h=\frac{(\epsilon-r^2)\psi^1+(\epsilon-r^1)\psi^2}{r^1r^2-\epsilon^2}
\end{equation*}
where
\begin{equation*}
\psi^1=-\frac{2(\eta^1+1)^\alpha(\eta^1-1)^\beta}{c_1}\displaystyle\int{(\eta^1+1)^\beta(\eta^1-1)^\alpha\eta^1\, d\eta^1},
\end{equation*}
\begin{equation*}
\psi^2=-\frac{2(\eta^2+1)^\beta(\eta^2-1)^\alpha}{c_1}\displaystyle\int{(\eta^2+1)^\alpha(\eta^2-1)^\beta\eta^2\, d\eta^2},
\end{equation*}
and
\begin{equation*}
\alpha=-\frac{c_1-c_2}{2c_1},\qquad \beta=-\frac{c_1+c_2}{2c_1}.
\end{equation*}
\bigskip

\bigskip

{\noindent} {\bf Separable case:}
$$
\xi^i(\eta^i)\ {\rm arbitrary}, \qquad \epsilon(\eta^1, \eta^2)=\phi_1(\eta^1)+\phi_2(\eta^2).
$$
The requirement of  flatness  of metric (\ref{g2}) leads to the following expressions for the functions $s_1(\eta^1), \ s_2(\eta^2)$ and $g_1(r^1, \eta^1), \ g_2(r^2, \eta^2)$:
$$
s_1=\frac{\phi_1^2c_2+2c_3\phi_1+2c_4}{2\phi'_1},\qquad s_2=-\frac{\phi_{2}^2c_2-2c_3\phi_2+2c_4}{2\phi'_2}, 
$$
$$
g_1=(c_1+c_3+c_2\phi_1)r^1,\quad g_2=-(c_1-c_3+c_2\phi_2)r^2,
$$
where $c_1,c_2,c_3$ and $c_4$ are arbitrary constants.  The Hamiltonian density is given by 
\begin{equation*}
h=\frac{(\epsilon-r^2)\psi^1+(\epsilon-r^1)\psi^2}{r^1r^2-\epsilon^2}
\end{equation*}
where 
$$
\psi^1=\displaystyle 2\sqrt{c_2\phi_1^2+2c_3\phi_1+2c_4}\, e^{K_1(\eta^1)}\int{\frac{\phi_1'e^{-K_1(\eta^1)}\xi_1}{\left(c_2\phi^2_1+2c_3\phi_1+2c_4\right)^{3/2}}\, d\eta^1},
$$
$$
\psi^2=\displaystyle -2\sqrt{c_2\phi_2^2-2c_3\phi_2+2c_4}\, e^{K_2(\eta^2)}\int{\frac{\phi_2'e^{-K_2(\eta^2)}\xi_2}{\left(c_2\phi^2_2-2c_3\phi_2+2c_4\right)^{3/2}}\, d\eta^2};
$$
here
\begin{equation*}
K_1(\eta^1)=\frac{c_1}{\sqrt{2c_2c_4-c_3^2}}\arctan{\left(\frac{c_2\phi_1+c_3}{\sqrt{2c_2c_4-c_3^2}}\right)},
\end{equation*}
\begin{equation*}
K_2(\eta^2)=\frac{c_1}{\sqrt{2c_2c_4-c_3^2}}\arctan{\left(\frac{c_2\phi_2-c_3}{\sqrt{2c_2c_4-c_3^2}}\right)}.
\end{equation*}
\bigskip

\bigskip
We expect that, in the  continuum limit,  the above formulae would provide Hamiltonian formulation of the full kinetic equation for dense soliton gas.

\section*{Acknowledgements}
We thank B. Doyon, G. El, P. Lorenzoni, M. Pavlov, and R. Vitolo for useful discussions.
PV's research was partially supported by GNFM of the Istituto Nazionale di Alta Matematica
(INdAM), the research project Mathematical Methods in Non-Linear Physics
(MMNLP) and by the Commissione Scientifica Nazionale -- Gruppo 4 -- Fisica Teorica
of the Istituto Nazionale di Fisica Nucleare (INFN) and PRIN 2017 \textquotedblleft Multiscale phenomena in Continuum
Mechanics: singular limits, off-equilibrium and transitions\textquotedblright,
project number 2017YBKNCE.  The research of EVF was supported by a grant from the Russian Science Foundation No. 21-11-00006, https://rscf.ru/project/21-11-00006/. 

\end{document}